\documentclass[%
 aip,
 amsmath,amssymb,
 reprint,%
]{revtex4-1}

\usepackage{upgreek}
\usepackage{graphicx}
\usepackage{dcolumn}
\usepackage{bm}

\usepackage[utf8]{inputenc}
\usepackage[T1]{fontenc}
\usepackage{mathptmx}
\usepackage{etoolbox}

\makeatletter
\def\@email#1#2{%
 \endgroup
 \patchcmd{\titleblock@produce}
  {\frontmatter@RRAPformat}
  {\frontmatter@RRAPformat{\produce@RRAP{*#1\href{mailto:#2}{#2}}}\frontmatter@RRAPformat}
  {}{}
}%
\makeatother
\begin{document}

\title{\textbf{A Cryogenic Penning Trap Based on Permanent Magnets}}%

\author{F. V\"olksen}
\affiliation{Heinrich Heine Universit\"at D\"usseldorf, D\"usseldorf, Germany}

\author{K. K. Anjum}
\affiliation{Heinrich Heine Universit\"at D\"usseldorf, D\"usseldorf, Germany}

\author{R. P. Czampiel}
\affiliation{Heinrich Heine Universit\"at D\"usseldorf, D\"usseldorf, Germany}

\author{N. Daehnhardt}
\affiliation{Heinrich Heine Universit\"at D\"usseldorf, D\"usseldorf, Germany}

\author{S. Gavranovic}
\affiliation{Institut für Physik, Johannes Gutenberg-Universit\"at, Mainz, Germany}

\author{J. Hoentges}
\affiliation{Heinrich Heine Universit\"at D\"usseldorf, D\"usseldorf, Germany}

\author{L. K\"urschner}
\affiliation{Heinrich Heine Universit\"at D\"usseldorf, D\"usseldorf, Germany}

\author{J. D. Regier}
\affiliation{Heinrich Heine Universit\"at D\"usseldorf, D\"usseldorf, Germany}

\author{D. Schweitzer}
\affiliation{Heinrich Heine Universit\"at D\"usseldorf, D\"usseldorf, Germany}
\affiliation{Institut für Physik, Johannes Gutenberg-Universit\"at, Mainz, Germany}

\author{P. Sch\"ops}
\affiliation{Heinrich Heine Universit\"at D\"usseldorf, D\"usseldorf, Germany}

\author{A. M. Thomas}
\affiliation{Heinrich Heine Universit\"at D\"usseldorf, D\"usseldorf, Germany}

\author{S. Wienhues}
\affiliation{Heinrich Heine Universit\"at D\"usseldorf, D\"usseldorf, Germany}

\author{B. Arndt}
\affiliation{Max-Planck-Institut für Kernphysik, Heidelberg, Germany}
\affiliation{GSI Helmholtzzentrum für Schwerionenforschung GmbH, Darmstadt, Germany}

\author{T. Imamura}
\affiliation{Institut f{\"u}r Quantenoptik, Leibniz Universität Hannover, Hannover, Germany}
\affiliation{Physikalisch-Technische Bundesanstalt, Braunschweig, Germany}

\author{S. Endoh}
\affiliation{RIKEN, Fundamental Symmetries Laboratory, Wako, Japan}
\affiliation{Graduate School of Arts and Sciences, University of Tokyo, Tokyo, Japan}

\author{B. M. Latacz}
\affiliation{CERN, Geneva, Switzerland}

\author{M. Leonhardt}
\affiliation{Heinrich Heine Universit\"at D\"usseldorf, D\"usseldorf, Germany}

\author{P. Micke}
\affiliation{Max-Planck-Institut für Kernphysik, Heidelberg, Germany}
\affiliation{GSI Helmholtzzentrum für Schwerionenforschung GmbH, Darmstadt, Germany}

\author{J. Morgner}
\affiliation{CERN, Geneva, Switzerland}

\author{D. Natakala}
\affiliation{Imperial College, London, UK}

\author{S. Stahl}
\affiliation{Max-Planck-Institut für Kernphysik, Heidelberg, Germany}

\author{H. Yildiz}
\affiliation{Institut für Physik, Johannes Gutenberg-Universit\"at, Mainz, Germany}

\author{K. Blaum}
\affiliation{Max-Planck-Institut für Kernphysik, Heidelberg, Germany}

\author{J. A. Devlin}
\affiliation{Imperial College, London, UK}

\author{Y. Matsuda}
\affiliation{Graduate School of Arts and Sciences, University of Tokyo, Tokyo, Japan}

\author{C. Ospelkaus}
\affiliation{Institut f{\"u}r Quantenoptik, Leibniz Universität Hannover, Hannover, Germany}
\affiliation{Physikalisch-Technische Bundesanstalt, Braunschweig, Germany}

\author{W. Quint}
\affiliation{GSI Helmholtzzentrum für Schwerionenforschung GmbH, Darmstadt, Germany}

\author{A. Soter}
\affiliation{Eidgen{\"o}ssisch Technische Hochschule Z{\"u}rich, Z{\"u}rich, Switzerland}

\author{J. Walz}
\affiliation{Institut für Physik, Johannes Gutenberg-Universit\"at, Mainz, Germany}
\affiliation{Helmholtz-Institut Mainz, Mainz, Germany}

\author{C. Smorra}
\affiliation{Heinrich Heine Universit\"at D\"usseldorf, D\"usseldorf, Germany}

\author{S. Ulmer}
\affiliation{Heinrich Heine Universit\"at D\"usseldorf, D\"usseldorf, Germany}
\affiliation{RIKEN, Fundamental Symmetries Laboratory, Wako, Japan}

\date{\today}

\begin{abstract}
We report on a Penning trap based on NdFeB permanent magnets operated at 4\,K with a magnetic field strength of 280\,mT. Reliable loading and confinement of protons, H$_2^+$ ions, and electrons was demonstrated with non-destructive single-particle detection sensitivity of  protons and H$_2^+$ ions using superconducting image-current circuits. The axial frequency of individual particles reaches a shot-to-shot stability of 47 parts-per-billion comparable to that of state-of-the-art precision Penning-trap experiments. Measurements of the proton modified-cyclotron frequency show a shot-to-shot scatter of $0.14$ parts per million (p.p.m.), presently limited by millikelvin-level temperature fluctuations of the permanent-magnet assembly. We outline a route towards improving this performance by more than an order of magnitude. This development offers broad potential for axial-mode-related precision measurements and cost-efficient Penning-trap experiments, and represents an important step towards compact, scalable, and transportable antiproton-trap systems.
\end{abstract}

\maketitle

\maketitle

\section{Introduction}
Cryogenic Penning traps \cite{brown1986geonium} are ultra-sensitive devices capable of non-destructively detecting single charged particles \cite{wineland1975principles} and measuring their characteristic oscillation frequencies over time scales extending a year~\cite{Bor22}. A variety of the most sensitive tests of the Standard Model of particle physics is based on Penning-trap techniques, including the most precise direct test of quantum electrodynamics through measurements of the magnetic moment of a single trapped electron~\cite{Fan23}. Penning traps furthermore enable ultra-precise tests of bound-state quantum electrodynamics based on highly charged ions~\cite{morgner2025testing}, atomic-mass measurements with parts-per-trillion precision~\cite{rainville2004ion} that provide input to neutrino mass constraints \cite{schweiger2024penning}, and to searches for exotic physics \cite{wilzewski2025nonlinear, smorra2019direct, budker2022millicharged}. Some recent developments focus on the implementation of quantum logic spectroscopy in Penning traps \cite{Cornejo2021}, and the use for quantum information science \cite{jain2024penning}. Beyond this type of precision spectroscopy, Penning traps play a pivotal role in the confinement of large-scale non-neutral plasmas, which is instrumental in the synthesis of antihydrogen from its constituent particles \cite{akbari2025be+}: positrons and antiprotons \cite{gabrielse1986prospects}.\\ 
The primary objective of our collaboration, the Baryon Antibaryon Symmetry Experiment (BASE), is to perform ultra-precise tests of the fundamental charge--parity--time (CPT) reversal invariance by comparing the fundamental properties of protons ($\mathrm{p}$) and antiprotons ($\bar{\mathrm{p}}$) with the highest possible precision. Using cryogenic Penning traps, we have compared the proton-to-antiproton charge-to-mass ratios, $q/m$, with a fractional precision of $16$ parts per trillion (p.p.t.)~\cite{Bor22}, measured the magnetic moment of the antiproton with a precision of $1.6$ parts per billion (p.p.b.)~\cite{Smo17}, and recently reported the first coherent spectroscopy of a single antiproton spin \cite{latacz2025coherent}. 
The systematic limitations encountered in our precision measurements at CERN~\cite{Dev19} motivated the recent development of the transportable Penning-trap system BASE-STEP~\cite{Smo23}. The central goal of this effort is to transport antiprotons out of the accelerator environment of CERN's antimatter factory into ultra-low-noise offline laboratories, where significantly improved experimental conditions can be realized. Major progress towards this goal has been demonstrated recently, through the successful transport of trapped protons~\cite{leonhardt2025proton} and antiprotons~\cite{Leonhardt2026RoadTransport}. Closely connected to this strategy is the development of cryogen-free permanent-magnet-based Penning traps that are substantially more compact, cost-effective, and simpler to operate than superconducting magnets. Such traps are insensitive to transient temperature spikes that occur during transport \cite{Smo23,leonhardt2025proton}, and consequently form compact and robust transport systems for exotic particles, such as antiprotons \cite{Smo15}, $\bar{\text{H}}_2^-$ \cite{myers2018cpt, schiller2026potential} or highly charged ions \cite{king2022optical}. 
\\
Several concepts of permanent-magnet-based Penning traps have been implemented, e.g.~a compact light ion trap \cite{Gom95}, a trap system for the spectroscopy and cooling of Be$^+$ and Ca$^+$ ions \cite{McMahon20, McMahon26}, or precursor traps for future precision measurements \cite{zhang2026development}, and other systems \cite{Sue02,Hoo15}. Similarly, electron-beam ion traps can be operated with permanent-magnet-based magnetic fields~\cite{Mic18,Mot00}, suitable for integrating Penning-trap systems. \\
In this paper, we report on the implementation of a compact cryogenic Penning-trap system based on NdFeB (Neodymium–Iron–Boron) permanent magnets \cite{herbst1991neodymium} assembled in an Aubert-configuration \cite{hugon2010design}. The cryogen-free system combines the permanent-magnet architecture with compensated Penning-trap electrodes \cite{Gab89}, and integrated cryogenic detection circuits \cite{nagahama2016highly}, enabling the long-term storage and non-destructive observation of charged particles. We describe the technical integration, the permanent-magnet assembly, and the trap system including the particle detectors for non-destructive frequency measurements. With this system we demonstrate loading of electrons, protons, and H$_2^+$-ions. We characterize for the first time a cryogenic permanent-magnet assembly based on cyclotron frequency measurements of trapped protons, and discuss the stability limits of the current setup. The manuscript concludes with a discussion of possible further improvement and future applications.

\section{Experimental Setup and Integration}
\begin{figure}
\centerline{\includegraphics[width=9.0cm,keepaspectratio]{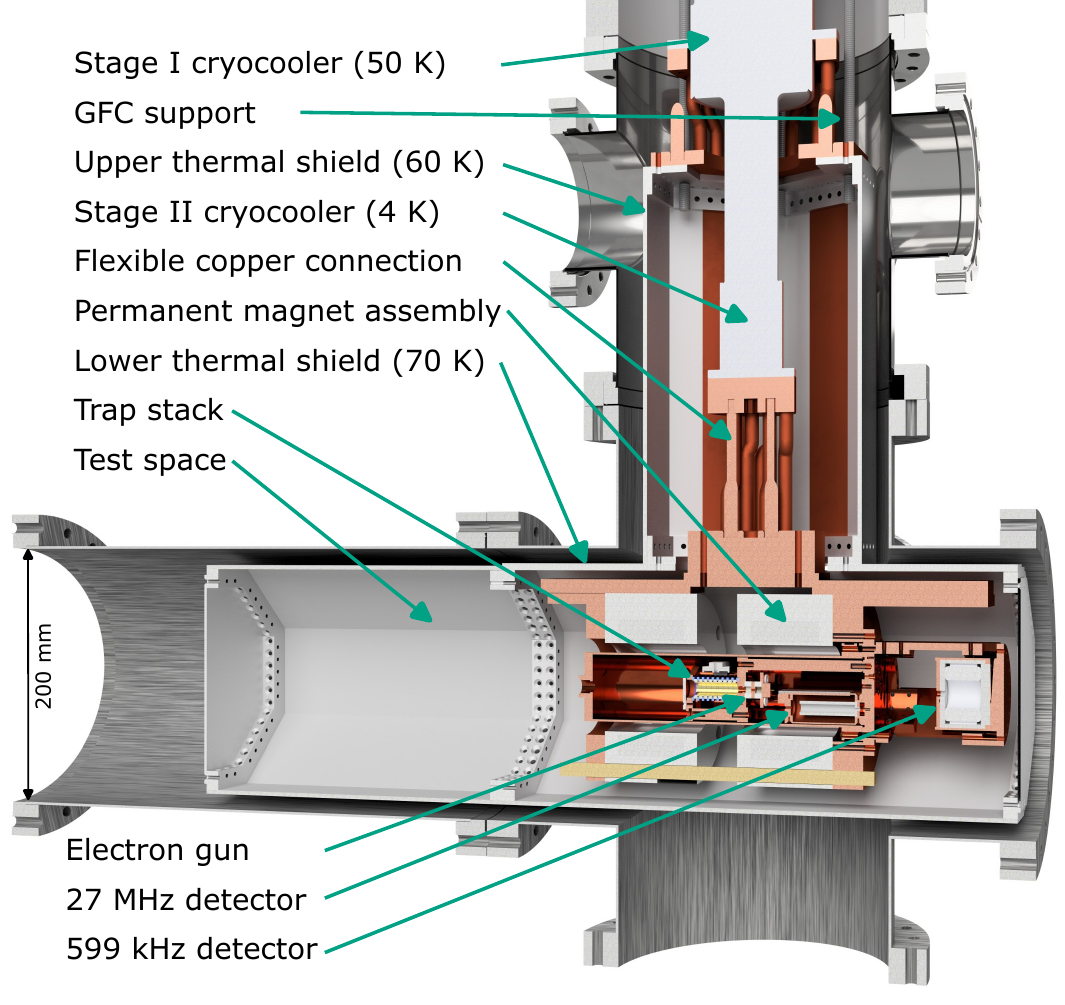}}
\caption{Cross-sectional view of the cryogenic permanent-magnet Penning-trap system. The setup consists of a compact cryogen-free cryostat with a two-stage Gifford--McMahon (GM) cryocooler, thermal radiation shields, and a permanent-magnet assembly in Aubert-configuration. The Penning-trap electrode stack is mounted in the center of the magnet assembly and coupled to two non-destructive image-current detection systems operating at approximately $599\,\mathrm{kHz}$ and $27\,\mathrm{MHz}$ for the measurement of  the axial frequency of stored ions and electrons, respectively. Flexible copper braids provide thermal anchoring to the cryocooler stages, while glass fiber composite (GFC) supports ensure mechanical stability with minimized thermal load. The system includes an electron gun for particle loading and a dedicated experimental test space for future integrations and applications.}
\label{fig:ExpSetup}
\end{figure}

\textbf{Overview:} A cross-sectional rendering of the cryogenic permanent-magnet Penning-trap system is shown in Fig$.\,$\ref{fig:ExpSetup}. The entire apparatus is integrated into a DN-200 six-way cross equipped with dedicated interface flanges for pressure and temperature monitoring, the routing of ultra-stable trap voltages, and the delivery of radio-frequency signals for particle manipulation and excitation to the trap. \\
The setup follows a compact cryogen-free design based on a two-stage Gifford--McMahon (GM) cryocooler, a Sumitomo RDE-412D4~\cite{SHI_RDE412D4}, providing a cooling power of $1.25\,\mathrm{W}$ at $4.2\,\mathrm{K}$ on stage II and up to $53$--$60\,\mathrm{W}$ at $\sim 43\,\mathrm{K}$ on stage I, with a typical compressor power consumption of approximately $7$--$8\,\mathrm{kW}$. \\
Inside the DN-200 vacuum chamber, a radiation shield made out of aluminum and copper is coupled to stage I of the GM cooler. 
The shield surrounds the trap stage, which is cooled by stage II of the cryocooler and consists of the permanent-magnet assembly, detectors for non-destructive particle detection \cite{nagahama2016highly} and Penning-trap electrodes \cite{Gab89} similar to the geometry used in \cite{Smo15}. 
The trap electrodes are mounted in a cryogenic, pre-pumped, pinched-off separate vacuum chamber in which -- due to cryopumping -- ultra-low pressures are achieved \cite{sellner2017improved}. Replacing the pinch-off with a cryogenic valve will be investigated in future setups. The cryogenic vacuum chamber sits in the Aubert magnet assembly using the trap support designed to align the central ring electrode precisely within the homogeneous-field region. 
Signals and voltages for the trap are guided from the outside into the outer vacuum chamber by standard feedthrough flanges shielded by aluminium boxes, which contain DC low-pass filter stages with a cut-off below 5\,kHz as well as radio-frequency switches. Inside the vacuum chamber filter boards are installed at stage I temperature. Signals are then guided to stage II filters mounted on the interface flange to the cryogenic vacuum chamber. Here, custom-made cryogenic feedthroughs based on Al$_2$O$_3$ provided by Kyocera Fine Ceramics guide the signals into the inner vacuum chamber via an indium-sealed vacuum flange made out of copper. \\
The entire assembly is mounted inside a 60$\,$cm wide, 80$\,$cm long and about 150$\,$cm high aluminum profile frame on wheels. The CF-200 vacuum chamber can be moved several centimeters in all directions with respect to the frame, allowing connection to different future setups, including beamlines for external injection.\\
\\
\textbf{Cryostat and Power Consumption:} The two cryo-stages -- the thermal shield as well as the trap-stage consisting of the compact trap-magnet-detection assembly -- are connected to the respective stages of the cold head by flexible connections. Stage I is connected with 16 oxygen free highly conductive (OFHC) copper ropes with 10\,mm diameter and $35 \,\text{mm}^2$ cross-section each, damping the influence of vibration of the pulses of the GM cooler on the trap and magnet. Stage II is connected to the trap stage with 7 parallel copper ropes with the same dimensions. Mechanical support is provided solely by 6 parallel 10\,mm wide glass fiber composite (GFC) tubes connecting the radiation shield to the upper end of the vacuum chamber and three GFC tubes between the radiation shield and the trap-stage. All tubes have a wall thickness of 1.5\,mm to minimize conductive heat transfer. \\
Once turned on, both stages reach thermal equilibrium in a cooling time of about 32$\,$h. 
The trap stage operates at 4.3\,K to 5.1\,K, and at the thermal shield at 40\,K to 70\,K, with the lowest temperatures at the heat exchanger interfaces of the cooler and the highest at the end of the radiation shield. We estimate a total heat load up to 25$\,$W at stage I, mainly due to thermal radiation, and a maximum of 78$\,$mW at the trap stage, consisting of 52$\,$mW radiation impact, 6$\,$mW from the mechanical support and 20$\,$mW from wiring and active electronics.
\\

\begin{figure}
\centerline{\includegraphics[width=9.0cm,keepaspectratio]{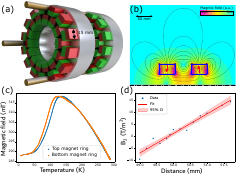}}
\caption{Permanent-magnet assembly and magnetic field characterization. 
(a) Schematic of the NdFeB ring-magnet assembly with polarity of the 56 individual magnets marked in red (north) and green (south). 
(b) Magnetic field lines obtained from an FEM calculation with the rotational axis indicated by the dashed line. 
(c) Measured magnetic field strength in the center of the magnet as a function of temperature of each individual magnet array ring. 
(d) Scaling of the magnetic bottle term $B_2$ as a function of magnet ring distance measured with a Hall probe at room temperature. A linear fit yields a variation $\Delta B_2 = 8.2 (1.4)$\,T/m$^2$ within a distance change of 1\,mm. The uncertainty results from scatter of the data.}
\label{fig:Magnet}
\end{figure}

\begin{figure}
\centerline{\includegraphics[width=9.0cm,keepaspectratio]{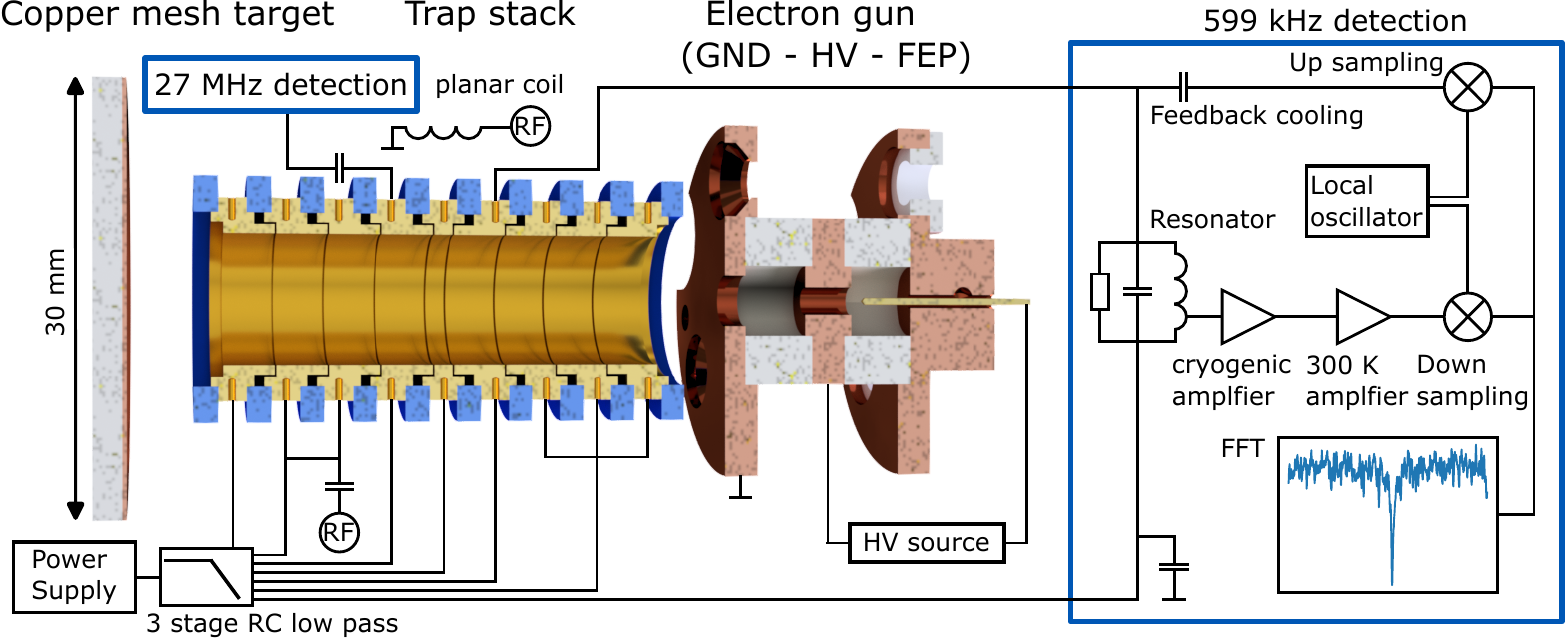}}
\caption{Electronics layout of the trap system. The trap is a 5-pole trap with extended endcaps, where sapphire rings provide electrical insulation and precise spacing between adjacent electrodes. The trap electrodes have an inner diameter of \(9\,\mathrm{mm}\), and their lengths are optimized for an orthogonalized and compensated Penning-trap design. Detection systems for the axial motion of ions at \(\nu_z \approx 599\,\mathrm{kHz}\) and the axial motion of electrons at \(\nu_z\approx 27\,\mathrm{MHz}\) with resonant circuits, cryogenic and 300\,K amplifiers and respective feedback lines are integrated. An etched tungsten field emission point (FEP) controlled by a high voltage (HV) electrode is used to load particles.}
\label{fig:TrapLayout}
\end{figure}

\textbf{Magnet:} Radial confinement in a Penning trap is given by the axial magnetic field $B(z)=B_0 + B_1 z + B_2 z^2+...$ at the position of the trapped particles. Here, the variation of the magnetic field along the axial direction is parametrized by the linear and quadratic coefficients $B_1$ and $B_2$. 
The assembly is shown in Fig.~\ref{fig:Magnet}\,(a) illustrating the mechanical construction and (b) showing a finite element method (FEM) 2.5\,D rotational symmetrical calculation of the magnetic field lines (black) and relative magnetic field strength on a linear scale. 
The Aubert magnet assembly \cite{Aubert1991, menzel2014design} as implemented here provides a strong $B_0$ with cylindrical symmetry, and a homogeneous region in the center, suitable for particle trapping. It consists of two radially magnetized annular structures. On one side of the central measurement region, the magnetization points radially inward and on the other side, it points radially outward.
The radial field components largely cancel near the center, while the axial components add. In practice, each ring is segmented into rectangular-shaped individual magnets made from NdFeB, a rare-earth permanent-magnet material based on the intermetallic compound $\mathrm{Nd_2Fe_{14}B}$ with 
typical remanence values 
around 1.3$\,$T \cite{herbst1991neodymium}.
 \\
For the individual magnets, we used commercially available N52 NdFeB permanent block magnets (MagnetMax GmbH) with dimensions of \(50.0 \times 15.0 \times 15.0\,\mathrm{mm}^3\), magnetized perpendicular to two opposing large surfaces. The individual NdFeB magnets are assembled inside aluminum support rings, each providing a machined structure that houses two concentric arrays of 14 block magnets, see Fig$.\,$\ref{fig:Magnet}\,(a). The two Al-support structures with the installed individual magnets are connected via three M10 threaded rods guided through precision boreholes, allowing adjustment of the axial spacing along the rotational axis $\vec{z}$ of the two magnet assemblies and therefore changing the quadratic magnetic field coefficient $B_2$. The magnet rings feature an inner diameter of \(64\,\mathrm{mm}\), compatible with the trap chamber dimensions, and an outer diameter of \(136\,\mathrm{mm}\).\\
For the assembly of the Aubert configuration, a total of 189 individual magnets were characterized using an F.W.~Bell Model 6010 Gauss/Teslameter, measuring the pole strengths at a defined distance of 32(1)\,mm, resulting in a median value of $B \approx 26.3\,\mathrm{mT}$. 
In the final assembly, we distributed the magnetization as symmetrically as possible between the two ring assemblies to minimize a magnetic gradient $B_1$ along the rotational axis. Magnets with $B \approx 26.3\,\mathrm{mT}$ were used for the inner assembly, while the outer rings employed alternating magnets with lower ($B \approx 26.1\,\mathrm{mT}$) and higher magnetization ($B \approx 26.5\,\mathrm{mT}$). Given the available components, the implemented assembly represents the best possible configuration regarding achievable field homogeneity.\\
Figure \ref{fig:Magnet}\,(c) shows results of a measurement in which the Hall probe was placed in the center of the magnet inside a room-temperature bore. The device was cooled to $\approx 7\,$K in a GM cryocooler driven test setup. At room temperature, a field strength of $B_0\approx287\,$mT was achieved. Upon cooling from room temperature, the magnetic field initially increases
because the remanent magnetization of NdFeB increases as thermal
spin fluctuations are suppressed. The broad maximum near $130\,\mathrm{K}$
is consistent with the turnover observed previously in such magnets, typically at temperatures of approximately \(130\)--\(150\,\mathrm{K}\)
\cite{Tanaka2006,He2018,Wolfers1996,Keavney1996}. 
Figure \ref{fig:Magnet}\,(d) shows the scaling of the magnetic bottle term $B_2$ at the trap center $z_0$ extracted from second order polynomial fits to on-axis magnetic field measurements with a Hall probe as a function of magnet distance at room temperature. A linear fit yields a variation $\Delta B_2 = 8.2 (1.4)$\,T/m$^2$ for a distance change $\Delta d = 1$\,mm, in excellent agreement with FEM calculation .
With geometrical alignment accuracy of $\approx250\,\upmu$m we demonstrated a homogeneity of $B_1=55(3)\,$mT/m and $B_2=-1.0(4)\,$T/m$^2$, measured with a Hall probe. Due to the residual pole-strength variation, the zero-crossings $z_1$ and $z_2$ of $B_1(z_1)=0$ and $B_2(z_2)=0$, respectively, are at different positions. A strategy to compensate for the residual $B_1$ is to ferro-shim the center, by placing soft-magnetic rings along the cylindrical axis asymmetrically shifted from the geometric center. For the cryogenic trap experiments described here, we deliberately tuned the system to $B_{2,t}\approx-30\,$T/m$^2$, later discussed in detail. \\
\\
\textbf{Trap Layout, Target, Electron Gun:} The trap setup is shown in Fig$.\,$\ref{fig:TrapLayout}. From left to right the assembly consists of a copper target, trap electrodes, and a field-emission electron gun, based on an etched tungsten tip. \\
The trap itself consists of cylindrical electrodes fabricated from oxygen-free electrolytic copper and coated with a \(7\,\upmu\mathrm{m}\) silver diffusion barrier followed by an \(8\,\upmu\mathrm{m}\) gold layer. The electrodes have an inner diameter of 9$\,$mm and are arranged in orthogonal and compensated design \cite{Gab89}.  This results in a quadrupolar-electric-field coefficient $C_2=18600/$m$^2$, similar to the layouts used in BASE \cite{Smo15}. The individual electrodes are spaced by sapphire rings, ideal due to its high thermal conductivity and low dielectric loss tangent. The electrodes are biased by a high-precision voltage source from Stahl Electronics, with a baseline voltage fluctuation of $\Delta V/V=10^{-8}$ \cite{StahlElectronicsUM}. 
Each DC line is guided to the trap through a sequence of RC filter stages located at room temperature and at the stage I and II of the cryocooler.
For particle manipulation, radio-frequency (rf) signals are guided to the trap via cryogenic coaxial cables, the line for axial dipole excitation is connected to an endcap electrode, while radial and quadrupolar sideband coupling drives are applied to a planar coil, mounted on the outside of the electrodes.\\
An ultra-sensitive image current detection system \cite{Nag16} based on a toroidal superconducting coil with an inductance of $2.4\,$mH, a free $Q$-value of 56\,400, and a free resonance frequency of 903$\,$kHz 
is implemented into the system. It is connected to the correction electrode on the electron gun side of the central ring electrode of the trap. The trap capacitance and connected electronics change its resonance frequency to $\approx599\,$kHz and the $Q$-value is lowered to $Q=32400$. In addition, a detection system at $\approx 27\,$MHz based on a solenoidal coil, similar to the one described in \cite{Ulm13}, with an inductance of $L \approx 2\,\mu$H, and a $Q$-value of 1050 is connected to the other correction electrode. The superconducting 599\,kHz detector 
is used for axial detection of ions, while the 27$\,$MHz detector is used for the detection of the axial frequency of electrons. \\
A detailed schematic detection layout is shown on the right hand side of Fig$.\,$\ref{fig:TrapLayout}. The $2.4\,$mH coil is tapped at a winding ratio $\approx1/5$. The tapped signal goes to an ultra low-noise amplifier based on a Sony 3SK164 GaAs field effect transistor. The output of the signal is amplified, down-converted, split, and analyzed by a Fast Fourier Transform (FFT) spectrum analyzer. The second branch of the down-converted signal is mixed-up with phase-shifted carrier, passes a voltage controlled attenuator and is fed back capacitively to the detector. This feedback loop allows us to adjust the electric temperature of the detection system and its effective parallel resistance $R_p=2\pi\nu L Q$ \cite{DUrso2003FeedbackOscillator}. 

\section{Preparation of Particle Clouds and Particle Cleaning}
To prepare clouds of trapped ions, the trap is initially biased to the power supply's maximum voltage of $-13.5\,$V and the electron gun is biased with $\approx800\,$V to initiate the electron beam. During loading, an electron current of $\approx 10\,$nA to $\approx 140\,$nA is applied at an electron beam energy of $\approx 42\,$eV. 
The electron current is typically kept on for about $30\,$s. This produces a cloud of trapped ions, defined by the electron energy and available elements in the trap usually consisting of protons, H$_2^+$ ions and different ions produced from the trap surfaces or residual gas. Depending on the species to be prepared, different cleaning procedures are applied.\\
The trap depth $V_0$ is adjusted such that the axial frequency
\begin{eqnarray}
\nu_z=\frac{1}{2\pi}\sqrt{\frac{2C_2qV_0}{m}}
\end{eqnarray}
of the desired species is tuned to resonance with the axial detection system and their axial mode is continuously cooled. Here, $q/m$ denotes the charge-to-mass ratio of the ion species and $C_2=18600/$m$^2$ is the quadrupole coefficient of the trap potential expansion $\phi(z,V_0)=V_0\sum_k C_kz^k$. Subsequently, a stored waveform inverse Fourier transform (SWIFT) excitation \cite{guan1996stored} is applied to the axial dipolar excitation electrode.  
The SWIFT is defined such, that all frequencies are excited, except for a $40\,$kHz band around the axial, modified cyclotron, and magnetron frequencies of the particles of interest as well as their respective sidebands \cite{Cor90}. Afterwards, the trap potential is rapidly ramped to a shallow depth of $V_0=500\,$mV. In this process, ions that gained energy from the SWIFT excitation escape from the trap, while the resistively cooled particles remain confined in the shallow potential well. After the evaporation step, the magnetron motion of the ions of interest is sideband-cooled by irradiating a quadrupolar drive at $\nu_z+\nu_-$. This transfers magnetron quanta to the axial mode, which is then cooled by the axial resonator \cite{Cor90}.
\\
\begin{figure}
\centerline{\includegraphics[width=9.0cm,keepaspectratio]{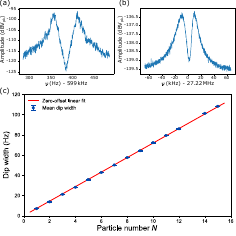}}
\caption{(a) Dip signal of 43(2) $H_2^+$ ions. (b) About 800 electrons loaded into the trap. (c) Dip width as a function of the number of trapped protons concluding to a single proton dip width of $\Delta\nu_z=7.11(2)\,$Hz.}
\label{fig:ParticleNumber}
\end{figure} 
For proton preparation, the same loading procedure is employed, but the cleaning is performed using high-voltage destabilization \cite{brown1986geonium}. In this method, the cloud of trapped protons is ramped into a deep axial potential, such that the radial electric force exceeds the Lorentz force for heavy ions, thereby destabilizing their radial confinement. This happens at $V_0>(qB_0^2)/(4mC_2)$. The trap parameters are chosen such that all ion species heavier than protons become unstable and are ejected from the trap, while protons remain confined. By following this procedure, and using the loading parameters given above, we typically load 50 to 100 particles per loading attempt. \\
For the loading of electrons, we invert the trap potential and activate the electron gun and the endcap on the far side of the electron gun biased with a negative voltage greater than the electron energy by about 10\,eV. 
With the active electron beam, we ramp the electron gun sided endcap up to negative voltages greater than the electron energy to accumulate electrons in the trap center. After electron-loading, negative ions are cleaned out of the trap by high-voltage destabilization, as used for protons. Fig$.\,$\ref{fig:ParticleNumber}\,(a) and (b) show clouds of 43(2) trapped H$_2^+$ ions and of order 800 electrons, respectively, after application of the procedures above. \\
After the preparation of a clean cloud of particles, a single particle is prepared, demonstrated with protons and H$_2^+$. 
To this end, we ground the endcaps and reduce the depth of the axial potential to evaporate particles. The resonator-thermalized axial mode results in a Boltzmann-distribution of energies. When lowering the axial trapping potential to thermal energies, hotter particles escape from the trap. After each evaporation step, the width of the axial dip is recorded. A two-parameter least-squares minimization correlates the measured dip width with particle number \cite{wineland1975principles}. For protons, the results of this particle reduction protocol are shown in Fig$.\,$\ref{fig:ParticleNumber}\,(c). The resulting axial single proton dip width is $\Delta\nu_z=7.11(2)\,$Hz.

\section{Trap Optimization}
\begin{figure}
\centerline{\includegraphics[width=9.0cm,keepaspectratio]{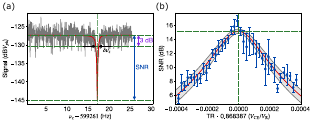}}
\caption{(a) Spectrum of a single-particle dip (grey). The signal-to-noise ratio (SNR), as well as the $3\, \mathrm{dB}$ dip width $\delta \nu_z$ based on a least-squares fit (red) are presented. (b) Dip SNR as a function of the tuning ratio (TR), defined as the ratio of voltages $V_\mathrm{CE}$ and $V_\mathrm{R}$, applied to the correction electrode and the ring electrode. From a least-square fit the optimum compensation TR at the maximum SNR is extracted.}
\label{fig:TROPT}
\end{figure}

To optimize the electrostatic axial trapping potential, with a single particle in the trap, the voltage applied to the correction electrodes $V_\text{CE}$ is scanned with a constant ring electrode voltage $V_\text{R}$ and the endcaps grounded, resulting the trap depth $V_0= V_\text{R}$. The voltage ratio $\text{TR}=V_\text{CE}/V_\text{R}$ is the tuning ratio. 
Changing this ratio affects the anharmonic contributions $C_4$ and $C_6$ of the trapping potential $\phi(z, V_\text{R})=V_\text{R}\sum_{k}C_{2k}z^{2k}$. In thermal equilibrium with the axial detection system, the particle shorts the thermal noise of the detector, as shown in Fig$.\,$\ref{fig:TROPT}\,(a), while executing a continuous random walk in axial energy space, where the energy is Boltzmann distributed with the probability density function $w(E_z)=(1/(k_\text{B}T_z))\exp\left(-E_z/(k_\text{B}T_z)\right)$. As a consequence, with the detector correlation time constant $\tau_z=(m/R_p)(D_\text{eff}/q)^2\approx30\,$ms being much shorter than typical FFT averaging times ($\geq16\,$s), we observe an average FFT spectrum with the particle's axial frequency 
\begin{eqnarray}
    \nu_z=\nu_{z,0}\left(1+\frac{3}{4}\frac{C_4}{C_2^2}\frac{E_z}{qV_\text{R}}+\frac{15}{16}\frac{C_6}{C_2^3}\left(\frac{E_z}{qV_\text{R}}\right)^2+...\right)
\end{eqnarray}
distributed over the thermal energy states \cite{brown1986geonium, Major2005ChargedParticleTraps}. In presence of anharmonic trap coefficients ($C_4, C_6, \cdots$), this leads to a reduction of the dip signal-to-noise ratio. When tuning the trap coefficients by changing the TR, the signal-to-noise ratio of the single-particle dip changes, reaching a maximum at the dynamical compensation point of the coefficient combination $\frac{\Delta\nu_z}{\nu_{z,0}}=\frac{3}{4}\frac{C_4}{C_2^2}\frac{E_z}{qV_\text{R}}+\frac{15}{16}\frac{C_6}{C_2^3}\left(\frac{E_z}{qV_\text{R}}\right)^2$. A result of such a tuning ratio scan is shown in Fig$.\,$\ref{fig:TROPT}\,(b),
where the tuning ratio was scanned in a range of $\pm0.0004$, reaching its maximum at the ideal tuning ratio. For axial frequency measurements at high resolution in short averaging times, we tune the trap to this optimum trap potential working point.

\section{Axial Frequency Measurements and Stability}

Combining a feedback-cooled resonator \cite{DUrso2003FeedbackOscillator} with operation at the optimal working point identified from a TR scan yields a signal-to-noise ratio of $\mathrm{SNR}=17.1(1)\,$dB and a single proton dip width of $\delta\nu_z\approx0.571(3)\,$Hz.
At these measurement parameters, the current best frequency resolution $\sigma\propto\Delta\nu_\text{z}/\text{SNR}$ is achieved. To characterize the stability of the axial oscillator, we record sequences of axial frequency dip spectra, averaged for $\tau_\text{avg}=90\,$s, and extract from each obtained spectrum the axial frequency $\nu_z$ of the particle by a least-squares fit of the known resonance line \cite{Win75} to the data. 
\begin{figure}
\centerline{\includegraphics[width=9.0cm,keepaspectratio]{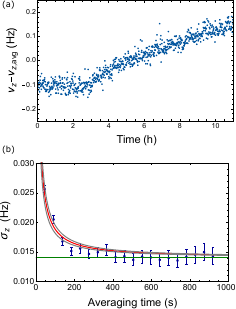}}
\caption{(a) Axial frequency $\nu_z (t)-\nu_{z,0}$ with $\nu_{z,0} \approx 599.274\,$kHz as a function of time. (b) Allan deviation of the frequency measurements. A frequency fluctuation noise floor of 15$\,$mHz is achieved, consistent with independently measured power supply noise (green line). The long term frequency drift of about 300\,mHz visible in (a) is in reasonable averaging times of no importance.}
\label{fig:Axial}
\end{figure} 
The measured data is displayed in Fig$.\,$\ref{fig:Axial}\,(a), where $\nu_z (t)-\nu_{z,0}$ is shown at $\nu_{z,0} \approx 599.274\,$kHz. The slow 300$\,$mHz peak-to-peak frequency drift that accumulates over about 12$\,$h corresponds to an equivalent voltage drift of $5\,\upmu$V, and is predominantly temperature correlated. Further stabilization is possible, but currently not of high relevance given the excellent short-term stability.\\ 
For a more detailed characterization, we evaluate the Allan deviation \cite{allan1966statistics} of the frequency sequence; the result is shown in Fig$.\,$\ref{fig:Axial}\,(b). Starting from a shot-to-shot frequency scatter of 28$\,$mHz, we observe a $\propto 1/\sqrt{\tau}$-scaling that converges after about 400$\,$s of averaging to the baseline axial frequency noise of $\approx15\,$mHz, which is the shot-noise-floor of the voltage source, independently measured by a reference multimeter.  
The $\propto 1/\sqrt{\tau}$-scaling is quantitatively understood and related to the SNR, currently limited by the residual anharmonicity of the trapping potential. For further optimization, a 7-electrode trap can be implemented, as demonstrated in \cite{bergstrom2002smiletrap, heisse2017high}. In any case, the axial frequency stability obtained in this trap is similar to our high-precision experiments at CERN \cite{Smo15}, and would allow for quantum transition spectroscopy of single electrons and combined with a stronger magnetic bottle even nuclear spins. Furthermore, it allows for phase-sensitive axial signal detection and self-excited oscillation measurements \cite{dur05} in future experiment campaigns.

\section{Cyclotron Frequency Measurements and Magnet Characterization}
\begin{figure}
\centerline{\includegraphics[width=9.0cm,keepaspectratio]{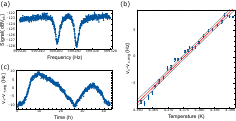}}
\caption{(a) Double dip signal used for cyclotron frequency measurements. The two dips arise from amplitude modulation of the axial oscillator while a sideband drive at $\nu_\text{rf}\approx\nu_+-\nu_z$ is applied. Sampling such spectra together with axial frequency measurements allows to extract the cyclotron frequency $\nu_c(t)$. (b): correlation of $\nu_c(t)-\nu_{c,0}$ and measured trap stage temperature. The linear correlation coefficient obtained from this measurement is 
$(d\Delta \nu_c/(dT \nu_c))=1.35(2)\cdot10^{-4}/$K. (c): cyclotron frequencies $\nu_c(t)-\nu_{c,0}$ with $\nu_{c,0}=4.269\,$MHz, measured within a time window of about 42$\,$h. The peak-to-peak fluctuation of the measurement series is 18$\,$Hz.}
\label{fig:MagFieldCorre}
\end{figure}

\textbf{Protocol:} To characterize the properties of the permanent-magnet assembly, we apply cyclotron frequency measurements using the sideband technique described in \cite{Bor22} and \cite{Cor90}. In short, we first measure the axial frequency $\nu_z$, and apply subsequently a quadrupolar sideband drive at $\nu_\text{rf}=\nu_+-\nu_z$. This continuously transfers axial-mode into modified-cyclotron-mode energy, which leads to modulation of the particle's axial motion amplitude. In the FFT spectrum of the amplitude-modulated axial oscillation, two signals at frequencies $\nu_l$ and $\nu_r$ appear, as shown in Fig$.\,$\ref{fig:MagFieldCorre}\,(a). The particle frequencies are obtained by least-squares fitting to the measured spectra, while $\nu_\text{rf}$ is defined by a calibrated frequency generator. From such measurements, the modified cyclotron frequency $\nu_+$ is obtained as \cite{Smo15}
\begin{eqnarray}
    \nu_+=\nu_\text{rf}+\nu_\text{l}+\nu_\text{r}-\nu_\text{z}.
\end{eqnarray}
The free cyclotron frequency $\nu_c=(q B_0)/(2\pi m_p)$ is then calculated by application of the invariance theorem $\nu_c^2=\nu_+^2+\nu_z^2+\nu_-^2$ \cite{gabrielse2009sideband}, where the magnetron frequency is obtained as $\nu_-\approx\nu_z^2/(2\nu_+)$.\\
We adjusted the magnet separation to deliberately apply a non-zero $B_2$ in the trap center.  This allows for application of the continuous Stern Gerlach effect \cite{Deh86}, axial temperature measurements, and read-out of radial mode energies. The downside is that $B_2$ leads to axial frequency walks during sideband coupling and loss of particle signal-to-noise ratio, which can be compensated by executing modified cyclotron sideband coupling with clouds of up to ten particles. In a next iteration an in-situ magnetic tuning system is planned to be implemented \cite{devlin2019superconducting}.\\
\\
\textbf{On-Axis Magnetic Field:} To characterize the properties of the magnetic field, we measure cyclotron frequencies $\nu_c$ as a function of axial particle position in the trap. To do that, we superimpose linear axial electric-field gradients over the trap center, by deliberately applying voltage asymmetries to the correction electrodes. The calibration of particle position as a function of offset voltage is extracted from potential calculations. 
\begin{figure}
\centerline{\includegraphics[width=9.0cm,keepaspectratio]{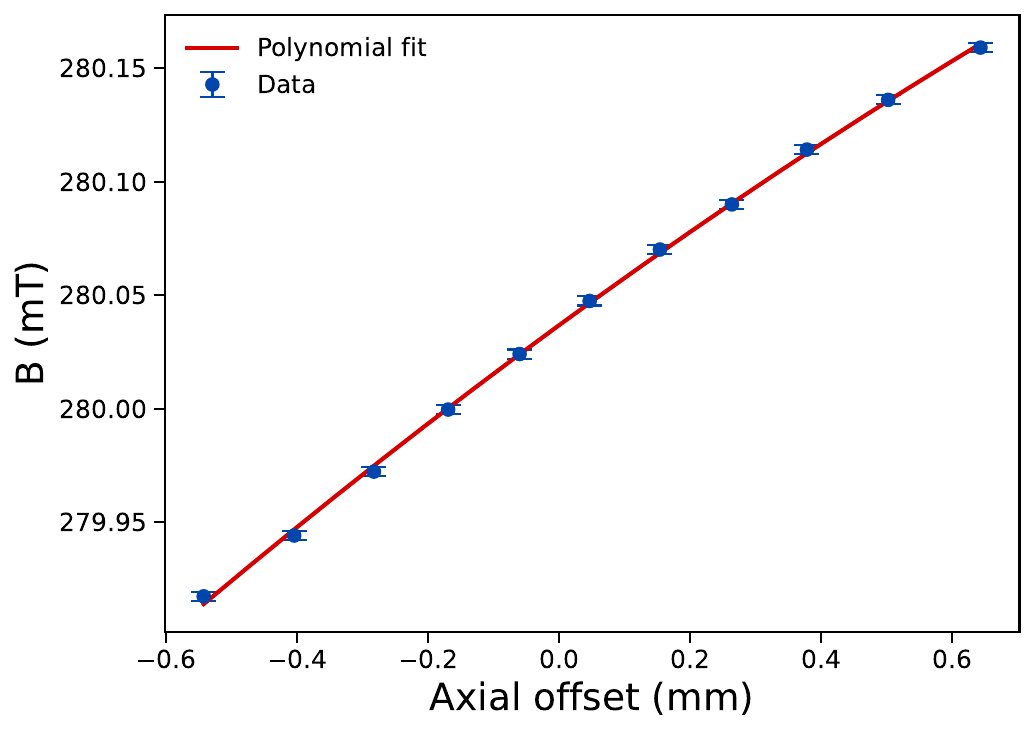}}
\caption{Trapped-particle magnetic field measurement. The axial position of the particle was changed by superimposing an electric-gradient field to the trap that shifts the trapped particles along the trap axis. At different axial positions, the cyclotron frequency has been measured. A parabolic fit to these data yields the magnetic gradient and magnetic bottle coefficients $B_1=211.0(1.8)$\,mT/m and $B_2=-28.4(5.1)$\,T/m$^2$}
\label{fig:MagFieldB1}
\end{figure} 
The results of measurements with a total particle shift of 1.2$\,$mm along the trap axis are shown in Fig$.\,$\ref{fig:MagFieldB1}. A parabolic fit to these data gives estimates of the magnetic gradient and magnetic bottle coefficients $B_1=211.0(1.8)$\,mT/m and $B_2=-28.4(5.1)$\,T/m$^2$ with uncertainties extracted from the fit. The magnetic gradient coefficient results from the $0.2\,\%$ field strength deviations of the poles of the individual magnets from their intended value, while the magnetic bottle term $B_2$ has been introduced deliberately in this run, by tuning the distances between the magnets. \\

\begin{figure}
\centerline{\includegraphics[width=9.0cm,keepaspectratio]{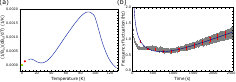}}
\caption{(a): Fractional temperature scaling $(1/B_0)(dB_0/dT)$ of the NdFeB 
magnets as a function of temperature. The solid blue line represents data measured with a Hall probe, the zero-crossing at $\approx125\,$K is due to a known spin-reorientation transition. The red data point represents the temperature scaling coefficient obtained by the cyclotron frequency measurement, the golden point is from \cite{adambukulam2021ultra}. (b): Allan deviation of measured data. The red points represent the Allan deviation of the measured cyclotron frequencies, the blue solid line is a fit to these data where the fitting function contains noise averaging $\propto 1/\sqrt{\tau}$ and temperature drift $\propto \tau$. The gray data is the Allan deviation of measured magnet temperatures scaled to frequency fluctuations, using the measured temperature scaling coefficient 
$d\nu_+/(d T) = 574.7(8.4)\,$Hz/K. The minimum achieved temperature fluctuation of about $1.1(1)\,$mK in the Allan deviation would correspond to a frequency fluctuation of $\Delta\nu_+=0.639(58)\,$Hz, in 1-sigma agreement with the best frequency fluctuation $\Delta\nu_+=0.614(31)\,$Hz shown in the Allan deviation.}
\label{fig:MagFieldADE}
\end{figure}

\textbf{Sideband coupling and axial temperature determination:} In the current setup, the magnet was deliberately tuned so that a quadratic magnetic component $B_2$ was superimposed on the trap center. This component couples radial motions like the modified cyclotron mode via their magnetic moment to the axial frequency \cite{brown1986geonium}. Energy $E_+$ in the cyclotron mode shifts the axial frequency $\nu_z$ by 
\begin{eqnarray}
    \Delta\nu_{z,+}(E_+)=\frac{1}{4\pi^2m_p\nu_{z,0}}\frac{B_2}{B_0}E_+.
\end{eqnarray}
Using the previously measured $B_2$ obtained by characterizing the magnetic field as a function of axial position, the temperature of the axial detection system can be determined. \\
To do this, we prepare a single particle, and measure axial frequency sequences $\nu_{z,i}$. Between neighboring axial frequency measurements we irradiate a sideband drive at frequency $\nu_{\text{rf}} \approx \nu_+-\nu_\text{z}$. 
\begin{figure}
\centerline{\includegraphics[width=9.0cm,keepaspectratio]{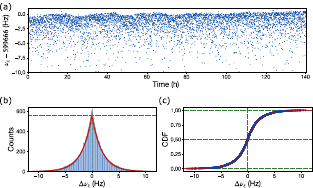}}
\caption{(a) Axial frequency as a function of time. Between neighboring measurements, a sideband drive at $\nu_\mathrm{rf} \approx \nu_+-\nu_z$ is applied for 20\,s, during which $E_+$ is coupled to the resonators Boltzmann distribution, performing a random walk. A clear baseline frequency $\nu_{z,0}$ emerges, slightly structured by laboratory temperature fluctuations. (b) Histogram of the frequency scatter $\Delta \nu_{z,i} = \nu_{z,i+1}- \nu_{z,i}$. The red line represents the probability density function of a Laplace distribution. (c) Empirical cumulative distribution function (CDF) of the scatter data, with a least-squares fit to the corresponding theoretical distribution function for $B_2T_z= -211.79(6) \, \mathrm{TK/m^2}$.}
\label{fig:MagFieldCorr_B2}
\end{figure} 
This couples the axial mode to the cyclotron mode, which leads to equilibration of the cyclotron mode energy with the axial detector such that $T_+=(\nu_+/\nu_z)T_z$. While the sideband drive is active, both coupled modes perform a random walk in the respective oscillator energy space, which freezes in the cyclotron mode to an energy value $E_{+,i}$ when the sideband drive is turned off. The imprinted radial magnetic moment related to the cyclotron energy $E_{+,i}$ leads to a shift of the axial frequency $\Delta\nu_{z,+}(E_{+,i})$. The axial frequencies obtained by this measurement sequence are shown in Fig.~\ref{fig:MagFieldCorr_B2}\,(a).
Here a clear baseline frequency $\nu_{z,0}$ is observed \cite{latacz2024orders}, slightly structured by laboratory temperature fluctuations. Fluctuating negative axial frequency shifts with respect to the baseline resemble the thermal Boltzmann distribution of a one-dimensional oscillator coupled to a negative $B_2$. By evaluating frequency differences $\Delta \nu_{z,i}=\nu_{z,i+1}-\nu_{z,i}$ and plotting the results in a histogram, the plot shown in Fig.~\ref{fig:MagFieldCorr_B2}\,(b) is obtained, which represents a two-sided Boltzmann distribution with a width $\sigma$ proportional to the product $B_2T_z$. Its probability density function is proportional to $p(\Delta \nu_z)\propto\exp\left(-\vert \Delta\nu_{z} \vert /(\alpha k_B T_z)\right)$, with $\alpha =  (d \Delta\nu_{z,+}/d E_+)(\nu_+/\nu_{z})$.
\\
Fig.~\ref{fig:MagFieldCorr_B2}\,(c) shows the normalized cumulative distribution function (CDF) of the measured results. Combining a least-squares fit to the corresponding CDF with the measured frequencies $\nu_z$ and $\nu_+$ and the previously determined values $B_0 = 280.005(1)$\,mT and $B_2=-28.4(5.1)$\,T/m$^2$ yields an axial detector temperature of $T_z=7.5(1.3)\,$K, consistent with the detector's signal-to-noise ratio at the chosen amplifier settings \cite{nagahama2016highly}. 
\\
\\
\textbf{Cyclotron Frequency Stability and Temperature Correlation:}
To characterize the stability of the cryogenic permanent-magnet assembly based on trapped-particle cyclotron frequency measurements, we record synchronized magnet temperature measurements and alternate measurements of axial $\nu_z$ and cyclotron-sideband $\nu_l$ and $\nu_r$ frequencies, to obtain a time series of $\nu_+$ frequencies. Each of the measured spectra is averaged for 90$\,$s while the thermometry is read out at a sampling rate of 1.5$\,$s. 
A representative result, measured for a time range of about 42$\,$h, is shown in Fig$.\,$\ref{fig:MagFieldCorre}\,(c). 
Here, the proton cyclotron frequency of about 4.269$\,$MHz varies within a peak-to-peak range of about 18$\,$Hz, while in the same time range the temperature of the magnet was changing by about $30\,$mK. 
A direct frequency to temperature correlation plot is shown in Fig$.\,$\ref{fig:MagFieldCorre}\,(b), with a dominantly linear correlation coefficient of 574.7(8.4)$\,$Hz/K, in relative units $(d\Delta\nu_c/(dT\nu_c))=(d\Delta B_0/(dT B_0))=1.35(2)\cdot10^{-4}/$K. As shown in Fig$.\,$\ref{fig:MagFieldADE}\,(a), this value is consistent with $(d\Delta B_0/(dT B_0))$ scalings measured with a Hall probe in a test setup. Here the blue solid line represents the Hall-probe measurements, while the red data point at 4.4$\,$K are results of trapped-particle-based $\nu_+$ measurements in the present setup. The second point at 30$\,$mK is from \cite{adambukulam2021ultra}, where a NdFeB Halbach assembly was characterized with an electron QUBIT in a dilution refrigerator. \\
To understand the measured stability limit, we evaluate the Allan deviation \cite{allan1966statistics} of the measured data, which is shown in Fig$.\,$\ref{fig:MagFieldADE}\,(b). 
The red points represent the Allan deviation of the measured cyclotron frequencies, the blue solid line is a fit to these data where the fitting function contains noise averaging $\propto A_1/\sqrt{\tau}$ and temperature drift $\propto A_2 \tau$. The parameter $A_1$ is determined by the properties of the particle dip, and the parameter $A_2$ by the temperature drift. 
The gray data is the Allan deviation of measured magnet temperatures scaled to frequency fluctuations, using the measured temperature scaling coefficient $(d\Delta B_0/(dT B_0))=1.35(2)\cdot10^{-4}/$K. The minimum achieved temperature fluctuation  of about $1.1(1)\,$mK in the Allan deviation would correspond to a frequency fluctuation of $\Delta\nu_+=0.639(58)\,$Hz, consistent with the lowest measured frequency fluctuation of $\Delta\nu_+=0.614(31)\,$Hz after about 700\,s averaging time indicated by the Allan deviation. This indicates, that the presently measured frequency stability is limited by an interplay of particle signal-to-noise ratio leading to a large parameter $A_1$, and temperature fluctuations of the trap stage.

\section{Improvements and Upgrades}
The presented characterization hints to straightforward first order improvements of the setup. The most immediate upgrades address detection sensitivity and magnetic-field stability. It is planned to install a self-shielding solenoid \cite{gabrielse1988self} around the cryogenic vacuum chamber. For homogeneous magnetic field disturbances shielding factors of $>200$ have been demonstrated with such a system \cite{Dev19}. Together with improved temperature stabilization consisting of an array of temperature sensors and local heating elements, these measures should improve the achievable cyclotron-frequency stability. Bringing all these upgrades together, we expect to reach future cyclotron-frequency stability of 10 parts per billion for subsequent frequency measurements with a $4\,$K-trap.\\
A dedicated in-situ magnetic field control system consisting of ferromagnetic shimming and superconducting shimming coils, as used in standard NMR magnets, will provide controlled adjustment of the leading magnetic-field coefficients. We aim to tune the $B_1$ gradient in a range of $\pm100\,$mT/(m$\cdot$A) and the $B_2$ coefficient within $\pm20\,$T/(m$^2\cdot$A). These improvements will be a permanent addition to the presented trap system.\\
Technological developments in this trap will focus on the characterization and optimization of components relevant for future precision experiments. In particular, cryogenic valves \cite{Leonhardt2026RoadTransport, Smo23, klimes2023cryogenic, sturm2019alphatrap} will enable particle injection and ejection while maintaining the cryogenic vacuum conditions required for long-term antiproton storage. Furthermore, SQUID-based image-current detectors \cite{weisskoff1988rf} can reduce the detector noise floor and particle temperature, thereby improving measurement precision. Both technologies will be integrated, optimized, and characterized in the presented setup with a view towards their implementation in future precision experiments.\\
Future permanent-magnet trap systems may be based either on miniaturized magnet arrays with reduced cooling requirements or on extended double-trap configurations combined with a solenoidal magnetic field and dedicated beam optics for particle transfer. Of particular interest is the development of a trap system suitable for antiproton transport. For this purpose, the present setup must be converted into an open trap configuration that allows controlled injection and ejection of antiprotons. Maintaining the required in-trap pressure during both stationary operation and transport additionally requires differential pumping to suppress residual-gas flow into the trap, following the concepts implemented in our superconducting transportable trap system \cite{Smo23,leonhardt2025proton}. A liquid-helium reservoir is further required to maintain the cryogenic conditions and ultra-high vacuum during transport, preventing the desorption of helium and hydrogen from the trap chamber \cite{Leonhardt2026RoadTransport}.\\
Finally, the integration of additional particle sources will extend the range of accessible species. In particular, sources for highly charged ions \cite{Mic18, rausch2026deceleration} and a dedicated positron source will enable first experiments with trapped exotic species. Together, these developments constitute important steps towards compact systems for the storage, manipulation, and transport of antimatter.

\section{Summary}
In summary, we demonstrate that a cryogenic NdFeB permanent-magnet system can support the complete experimental chain of a modern Penning-trap experiment, from particle production and species-selective preparation to long-term confinement and non-destructive single-particle frequency measurements. Using the trapped particles as in-situ magnetic-field probes, we characterize magnetic field coefficients and the cryogenic temperature coefficient and identify temperature fluctuations as the dominant limitation of the present cyclotron-frequency stability.\\
These limitations are technical rather than fundamental and can be substantially reduced through improved thermal stabilization, magnetic shielding, field correction, and higher-sensitivity detection. The combination of single-particle control, compact dimensions, and modest infrastructure requirements makes permanent-magnet Penning traps particularly attractive for dedicated experiments and transportable particle reservoirs. Our results thus establish a path toward compact, deployable systems for the storage, manipulation, and transport of antiprotons and other exotic particles, complementing high-precision experiments based on superconducting magnets.

\begin{acknowledgments}
We acknowledge advise from Peter Bl\"umler (Univ. Mainz) during construction of the permanent-magnet assembly. We are in particular thankful for the support by Heinrich Heine University Düsseldorf and RIKEN. We acknowledge financial support by Heinrich Heine University Düsseldorf, EIN Quantum NRW, DFG through the New Instrumentation Program (Grant No.~558683350), RIKEN, the Max-Planck Society, and the Max Planck, RIKEN, PTB Center for Time, Constants, and Fundamental Symmetries (C-TCFS). \\
\\
\end{acknowledgments}

\textbf{Author Contributions:} The project was initiated by C.S. and S.U. The experiment was conceptualized by F.V., C.S., and S.U. F.V. led the technical design and overall integration, with contributions from L.K., N.D., J.H., J.R., K.A., C.S., and S.U. P.S., F.V., S.W., and S.U. integrated the detection systems. S.G., D.S., C.S., F.V., and J.R. conceptualized and implemented the magnet system. J.H., F.V., and S.U. developed the electron gun. F.V., N.D., J.H., J.R., and S.W. developed and integrated the electronics. K.A., F.V., C.S., and S.U. developed the experiment control code. F.V., K.A., R.C., C.S., and S.U. acquired and analyzed the data.
\\
S.U., F.V., K.A., R.C., and C.S. wrote the manuscript. All authors contributed to the interpretation and discussion of the results and to the revision of the manuscript.
\\
Funding: S.U. (lead), C.S., and K.B.
\\

\textbf{Conflict of interest:}
The authors declare no competing interests.
\\
\\
\textbf{Data availability:} The data that support the findings of this study are available from the corresponding author upon reasonable request.
\\
\\
This article has been submitted to Review of Scientific Instruments. After it is published, a link will be included. Copyright (2026) BASE-Collaboration and Frederik V\"{o}lksen. This article is distributed under a Creative Commons Attribution-Non-Commercial-No-Derivs 4.0 International (CC-BY-NC-ND) License

\bibliographystyle{aipnum4-1-RSI.bst}
\bibliography{apssamp.bib}
\end{document}